%% file: main.tex
\documentclass[10pt,conference]{IEEEtran}
\IEEEoverridecommandlockouts
\usepackage{cite}
\usepackage{amsmath,amssymb,amsfonts}
\usepackage{algorithm}
\usepackage{algorithmic}
\usepackage{graphicx}
\usepackage{textcomp}
\usepackage{xcolor}
\usepackage{mathtools}
\usepackage{qcircuit}
\usepackage{braket}
\usepackage{svg}
\usepackage{verbatim}
\usepackage{hyperref}
\hypersetup{%
colorlinks=true,
citecolor=teal,
}
\def\BibTeX{{\rm B\kern-.05em{\sc i\kern-.025em b}\kern-.08em
    T\kern-.1667em\lower.7ex\hbox{E}\kern-.125emX}}
\begin{document}

\renewcommand{\sectionautorefname}{Section}
\renewcommand{\subsectionautorefname}{Subsection}
\newcommand{\algorithmautorefname}{Algorithm}

\title{Leveraging Zero-Level Distillation to Generate High-Fidelity Magic States
\thanks{This work has been submitted to the IEEE for possible publication. Copyright may be transferred without notice, after which this version may no longer be accessible.}}

\author{
\IEEEauthorblockN{Yutaka Hirano\IEEEauthorrefmark{1}, Tomohiro Itogawa\IEEEauthorrefmark{1}, and Keisuke Fujii\IEEEauthorrefmark{1}\IEEEauthorrefmark{2}\IEEEauthorrefmark{3}\IEEEauthorrefmark{4} 
}

\IEEEauthorblockA{
\IEEEauthorrefmark{1}\textit{Graduate School of Engineering Science}, \textit{Osaka University},\\
1-3 Machikaneyama, Toyonaka, Osaka 560-8531, Japan}

\IEEEauthorblockA{
\IEEEauthorrefmark{2}\textit{School of Engineering Science}, \textit{Osaka University},\\
1-3 Machikaneyama, Toyonaka, Osaka 560-8531, Japan}

\IEEEauthorblockA{
\IEEEauthorrefmark{3}\textit{Center for Quantum Information and Quantum Biology}, \textit{Osaka University},\\
1-2 Machikaneyama, Toyonaka 560-0043, Japan}

\IEEEauthorblockA{
\IEEEauthorrefmark{4}\textit{RIKEN Center for Quantum Computing (RQC)},
Hirosawa 2-1, Wako, Saitama 351-0198, Japan}

u965281c@ecs.osaka-u.ac.jp, u348506d@ecs.osaka-u.ac.jp, fujii@qc.ee.es.osaka-u.ac.jp
}

\maketitle

\begin{abstract}
\input{0-abstract}
\end{abstract}


\input{1-introduction}
\input{2-preliminary}
\input{3-main-proposal}
\input{4-performance-evaluation}
\input{5-conclusion}

\section*{Acknowledgment}
This work is supported by MEXT Quantum Leap Flagship Program (MEXT Q-LEAP) Grant No. JP- MXS0118067394 and JPMXS0120319794, JST COI- NEXT Grant No. JPMJPF2014, and JST Moonshot R\&D Grant No. JPMJMS2061.

\bibliographystyle{IEEETran}
\bibliography{refs}

\end{document}

%% file: 0-abstract.tex
Magic state distillation plays an important role in universal fault-tolerant quantum computing, and its overhead is one of the major obstacles to realizing fault-tolerant quantum computers.
Hence, many studies have been conducted to reduce this overhead.
Among these, Litinski has provided a concrete assessment of resource-efficient distillation protocol implementations on the rotated surface code.
On the other hand, recently, Itogawa~\textit{et al.} have proposed zero-level distillation, a distillation protocol offering very small spatial and temporal overhead to generate relatively low-fidelity magic states.
While zero-level distillation offers preferable spatial and temporal overhead, it cannot directly generate high-fidelity magic states since it only reduces the logical error rate of the magic state quadratically.
In this study, we evaluate the spatial and temporal overhead of two-level distillation implementations generating relatively high-fidelity magic states, including ones incorporating zero-level distillation.
To this end, we introduce (0+1)-level distillation, a two-level distillation protocol which combines zero-level distillation and the 15-to-1 distillation protocol.
We refine the second-level 15-to-1 implementation in it to capitalize on the small footprint of zero-level distillation.
Under conditions of a physical error probability of $p_{\mathrm{phys}} = 10^{-4}$ ($10^{-3}$) and targeting an error rate for the magic state within $[5 \times 10^{-17}, 10^{-11}]$ ($[5 \times 10^{-11}, 10^{-8}]$), (0+1)-level distillation reduces the spatiotemporal overhead by more than 63\% (61\%) compared to the (15-to-1)$\times$(15-to-1) protocol and more than 43\% (44\%) compared to the (15-to-1)$\times$(20-to-4) protocol, offering a substantial efficiency gain over the traditional protocols.

%% file: 1-introduction.tex
\section{Introduction}

Quantum computers promise to solve interesting problems that are also classically intractable, such as integer factoring~\cite{Shor1994Factoring}, quantum chemistry~\cite{aspuru2005simulated}, and solving large-scale linear systems~\cite{harrow2009quantum}.
Currently available quantum computers, also known as noisy intermediate-scale quantum computers (NISQ)~\cite{preskill2018quantum}, are too vulnerable to hardware errors to run complex quantum programs.
In order to realize theoretically proven quantum speedup, 
we need fault-tolerant quantum computers (FTQC).

FTQC is built on top of quantum error correcting (QEC) codes to suppress hardware errors, where computation is protected from errors by encoding logical qubits using many physical qubits, thereby adding spatial and temporal overhead.
Reducing the overhead, which is the main obstacle to realizing FTQC, is one of the most important issues in this field.
Consequently, various QEC codes are being explored and FTQC architectures are being designed~\cite{fujii2015quantum}.
Among these, the surface codes~\cite{Kitaev2003, Bravyi1998} are actively developed because of their relatively high error threshold with nearest-neighbor connectivity, and hence, we focus on them in this study.

Implementing Clifford gates on the surface codes in a fault-tolerant manner can be done relatively efficiently by using the lattice surgery~\cite{Horsman2012,Litinski2019GameOfSurfaceCodes}.
However, Clifford gates are not enough to form a universal gate set, with which we can run an arbitrary quantum algorithm; 
a non-Clifford gate is essentially required.
Protecting non-Clifford gates from errors is not trivial and hence a special gate, $T$ gate, is implemented via quantum gate teleportation with a special resource state known as a \textit{magic state}.
Instead of protecting $T$ gate directly, a high fidelity magic state is generated by a family of protocols called \textit{magic state distillation}~\cite{Bravyi2005}.
The spatial and temporal overhead of magic state distillation is one of the major bottlenecks of FTQC, especially when many $T$ gates are performed simultaneously, and hence, many studies have been conducted to reduce the overhead~\cite{Campbell2017Unified,Gidney2019EfficientMagicState,Litinski2019GameOfSurfaceCodes,Litinski2019magicstate,Herr2017LatticeSurgery, Itogawa2024Efficient}.

One primary reason why magic state distillation protocols require large spatial and temporal overhead is that they assume ideal Clifford gates, which requires the distillation process to run on logical qubits instead of physical qubits.
To mitigate this issue, Litinski~\cite{Litinski2019magicstate} has proposed resource-efficient implementations of magic state distillation protocols on the rotated surface code, accompanied by a detailed performance assessment.
The implementations reduce the overhead by carefully relaxing the ideality condition of Clifford gates to use fewer number of physical qubits per logical qubit.
Recently, Itogawa~\textit{et al.}~\cite{Itogawa2024Efficient} have proposed zero-level distillation, another efficient implementation of a magic state distillation protocol.
It runs distillation fault-tolerantly on physical qubits rather than logical qubits with carefully designing the error detection circuit, which essentially reduces the number of physical qubits required for distillation.
While zero-level distillation offers preferable spatial and temporal overhead, it cannot directly generate high-fidelity magic states since it only reduces the logical error rate of the magic state quadratically.
Therefore it is still not well understood how it contributes to reduce the spatial and temporal overhead for achieving a $T$ gate of a given target fidelity and parallelism.

In this study, we evaluate the spatial and temporal overhead of two-level distillation implementations generating relatively high-fidelity magic states, including ones incorporating zero-level distillation.
To this end, we introduce (0+1)-level distillation, a two-level distillation protocol which combines zero-level distillation and the 15-to-1 distillation protocol.
We refine the second-level 15-to-1 implementation in it to capitalize on the small footprint of zero-level distillation.
We also address the drawback of zero-level distillation, its relatively higher failure probability.
To assess the performance of these distillation implementations, we conduct two performance analyses.
One is assessing the spatial and temporal overhead of magic state distillation.
The other is a resource estimation of a Hamiltonian simulation, aiming to assess the impact of different distillation protocols on the overall performance of quantum computation.

The first analysis reveals that, under conditions of a physical error probability of $p_{\mathrm{phys}} = 10^{-4}$ ($10^{-3}$) and targeting an error rate for the magic state within $[5 \times 10^{-17}, 10^{-11}]$ ($[5 \times 10^{-11}, 10^{-8}]$), (0+1)-level distillation reduces the spatiotemporal overhead by more than 63\% (61\%) compared to the two-level 15-to-1 implementation and more than 43\% (44\%) compared to the two-level (15-to-1)$\times$(20-to-4) implementation, offering a substantial efficiency gain over the traditional protocols.
The second analysis reveals that, with $p_{\mathrm{phys}} = 10^{-4}$ $(10^{-3})$, adopting (0+1)-level distillation reduces the total spatial cost by 47\% (41\%) compared to using the (15-to-1)$\times$(15-to-1) protocol, and by 44\% (33\%) compared to the (15-to-1)$\times$(20-to-4) protocol, suggesting that the impact on the overall performance is substantial.

The rest of this paper is organized as follows.
In \autoref{sec:preliminary}, we provide basic definitions, notations, and existing distillation protocols, which are essential for understanding our proposal.
In \autoref{sec:0+1-level distillation}, we detail our proposal, (0+1)-level distillation.
In \autoref{sec:performance evaluation}, we conduct a comprehensive performance evaluation of (0+1)-level distillation.
Finally, \autoref{sec:conclusion} concludes the paper with a summary of our findings.

%% file: 2-preliminary.tex
\section{Preliminary}
\label{sec:preliminary}
In this section, we describe the existing magic state distillation protocols which we use in this paper, along with definitions and notations.
Specifically, we describe the single-level 15-to-1 implementation~\cite{Litinski2019magicstate} in \autoref{subsec:preliminary-single-15-to-1}, the two-level (15-to-1)$\times$(15-to-1) implementation in \autoref{subsec:preliminary-double-15-to-1}, and zero-level distillation in \autoref{subsec:zero-level-distillation}.

\subsection{Definitions and notations}
Let us begin by defining important classes of operators because they play important roles in the discussions that follow.
\textit{Pauli operators} are defined as one-qubit Pauli operators ($I$, $X$, $Y$, and $Z$) and their tensor products, together with factors $\pm1$ and ${\pm}i$.
Pauli operators acting on $n$ qubits, $\mathcal{P}_n$, form a group.
Clifford operators are defined as unitary operators that map a Pauli operator to a Pauli operator through conjugation.
\[
\mathcal{C}_n \coloneqq \{U \mid U\mathcal{P}_nU^{\dagger} \subset \mathcal{P}_n \}.
\]

Let us define a Pauli rotation along a Pauli operator $P$ with a rotation angle $\theta$,
\[
e^{i \theta P} = \cos \theta I + i \sin \theta P.
\]
For example, $S = e^{-\frac{i \pi}{4}Z}$ and $T = e^{-\frac{i \pi}{8}Z}$, up to global phase.
$e^{i \theta P}$ is a Pauli operator if and only if $\theta$ is a multiple of $\frac{\pi}{2}$, and it is a Clifford operator if and only if $\theta$ is a multiple of $\frac{\pi}{4}$.
Let us also define \textit{Pauli measurement} $M_P$ as $I + (-1)^{m}P$ where $P$ is a Pauli operator and $m \in \{0, 1\}$ is the measurement result.

\begin{figure}[t!]
\centering
\includesvg[width=8cm]{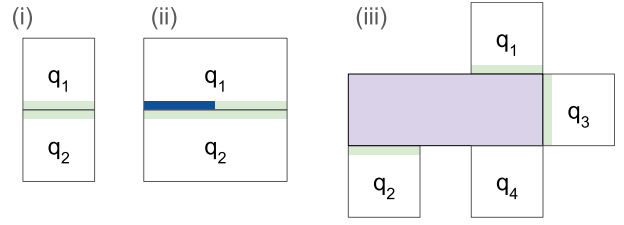}
\caption{Block diagrams depicting (i) $M_{ZZ}$, (ii) $M_{YZ}$, and (iii) $M_{ZZZ}$, respectively.
         Blue and green lines on the boundary of qubits represent logical operators involved in the lattice surgery operation, and the purple rectangle is used to connect qubits.} 
\label{fig:block diagrams}
\end{figure}

We use the rotated surface code as the error correcting code, and we use the time required to perform an error syndrome measurement (\textit{cycle}) as the time unit.
We call physical qubits which form one logical qubit a $d_X{\times}d_Z$ \textit{patch} where $d_X$ and  $d_Z$ are $X$ and $Z$ code distances, up to half of which the surface code tolerate $X$ and $Z$ errors, respectively.
We employ the lattice surgery~\cite{Horsman2012} to implement fault-tolerant logical Clifford gates~\cite{Litinski2019GameOfSurfaceCodes}.
In this paper, we illustrate quantum computation processes using two graphic representations.
One is a conventional quantum circuit, such as \autoref{fig:faulty-t-measurement}~(left).
The other is a block diagram, which contains logical qubit layout and lattice surgery information.
For example, \autoref{fig:block diagrams}~(i) depicts $M_{ZZ}$ where the green lines between the qubits signify logical $Z$ operators on the qubits,
(ii) illustrates $M_{YZ}$ with the blue line indicating a logical $X$ operator, and
(iii) represents $M_{ZZZ}$ where the purple rectangle connects qubits involved in the measurement ($q_4$ is not involved in the measurement).
Each of these operations takes $d$ cycles where $d$ is the code distance.

\subsection{15-to-1 protocol implementation}
\label{subsec:preliminary-single-15-to-1}
The 15-to-1 protocol implementation in~\cite{Litinski2019magicstate} employs a sequence of $\frac{\pi}{8}$ Pauli rotations on five logical qubits~\cite{Campbell2017Unified}.
Let $\Gamma$ be a set of Pauli operators defined as follows.
\[
 \Gamma \coloneqq \{\prod_{i \in s} Z_i \mid s \subset \{1, 2, 3, 4, 5\}, |s| \text{ is odd}\}.
\]
The product of $\frac{\pi}{8}$ Pauli rotations across $\Gamma$ is an identity operator acting on the five qubits.
\[
 \prod_{P \in \Gamma}e^{\frac{i\pi}{8}P} = I.
\]
Note that all the operators in $\Gamma$ commute and so do the $\frac{\pi}{8}$ Pauli rotations across $\Gamma$.
Because $Z_1 \in \Gamma$, 
\[
 \prod_{P \in \Gamma, P \neq Z_1}e^{\frac{i\pi}{8}P} = e^{-\frac{i\pi}{8}Z_1}.
\]
Thus we can compute $\ket{T} \coloneqq T\ket{+} = e^{-\frac{i\pi}{8}Z}\ket{+}$.
\[
\left(\prod_{P \in \Gamma, P \neq Z_1}e^{\frac{i\pi}{8}P}\right){\ket{+}}^{\otimes 5} = \ket{T} \otimes {\ket{+}}^{\otimes 4}.
\]
Because the rotations act on the five qubits non-trivially, we can detect errors on the rotations by measuring the last four qubits in the $X$ basis.

\begin{figure}[t!]
\begin{minipage}{5.5cm}
 \centering
 \[
 \Qcircuit @C=.5em @R=.6em {
   \lstick{\ket{\psi}} & /\qw & \multigate{1}{M_{PZ}}   & \qw                 & \qw              & \gate{P} & \qw      &    \rstick{e^{\frac{i\pi}{8} P}\ket{\psi}} \\
   \lstick{\ket{+}}    & \qw    & \ghost{M_{PZ}}        & \gate{X}            & \gate{T^\dagger} & \gate{M_X} \cwx[-1] &                                   \\
                       &        & \control[-1] \cwx[-1] & \cctrl{-1} \cwx[-1] &                                                                     \\
   }
 \]
\end{minipage}
\hfill
\begin{minipage}{1.0cm}
 \centering
 \includesvg[width=1.0cm]{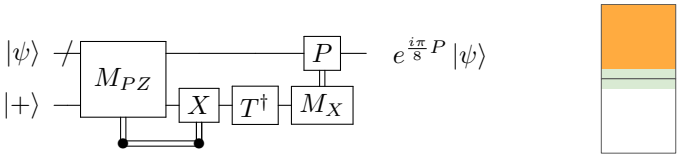}
\end{minipage}
\caption{Circuit of $\frac{\pi}{8}$ rotation along $P$ with a faulty $T^\dagger$ measurement (left) and its block diagram representation (right, $P = Z$).}
\label{fig:faulty-t-measurement}
\end{figure}

\begin{figure}[t!]
\centering
\includesvg[width=8.5cm]{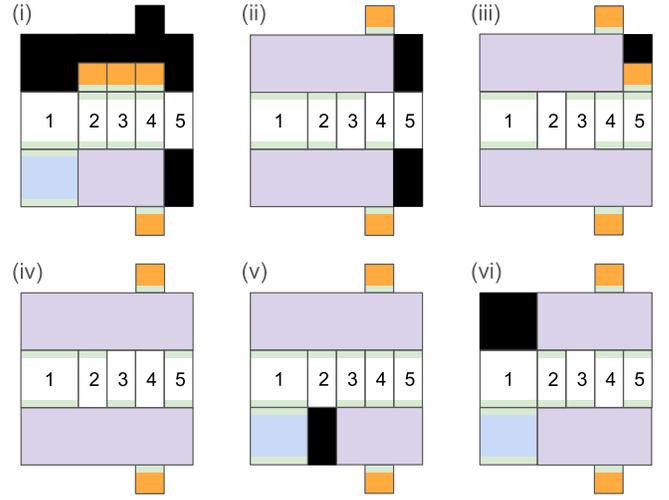}
\caption{Single level 15-to-1 protocol implementation consisting of six rounds.
         Numbered cells represent data qubits.
         Blue cells represent the magic state which is accessible to the outside of the distillation factory.
         Black cells represent unused qubits.}
\label{fig:single-level-15-to-1-implementation}
\end{figure}

The single-level 15-to-1 implementation uses \textit{faulty $T^\dagger$ measurement}, a faulty $T^\dagger$ operator followed by an $X$ measurement, to perform a $\frac{\pi}{8}$ rotation along a Pauli operator $P$, as illustrated in \autoref{fig:faulty-t-measurement}~(left).
We denote such a rotation as an orange cell connected with target qubits in block diagrams, as depicted in \autoref{fig:faulty-t-measurement}~(right).
The operation has an error rate comparable to the physical error rate.
The distillation process, which consists of six rounds, is depicted in \autoref{fig:single-level-15-to-1-implementation}.
For example, in the first round, $Z_2$, $Z_3$, $Z_4$, and $Z_2Z_3Z_4$ rotations are performed.
Since each round consists of $\frac{\pi}{8}$ rotations performed simultaneously, it completes in the time required to perform one $\frac{\pi}{8}$ rotation.
For rotations beyond single-qubit rotations (rotations along $Z_2$, $Z_3$, $Z_4$, and $Z_5$), the involved qubits and the qubits on which the faulty $T^\dagger$ gate is performed need to be connected, similarly to \autoref{fig:block diagrams}~(iii).
The purple regions in \autoref{fig:single-level-15-to-1-implementation} serve the purpose.
Let us call the regions \textit{ancilla regions}.
There are at most two ancilla regions for each round, namely the top ancilla region and bottom ancilla region.
At the end of the final round, qubits 2--5 are measured in the $X$ basis and the measurement results are checked to detect errors.
When an error is detected, we need to discard the quantum state at qubit 1.
We call this event \textit{distillation failure}.

\begin{figure}[t!]
\centering
\includesvg[width=5cm]{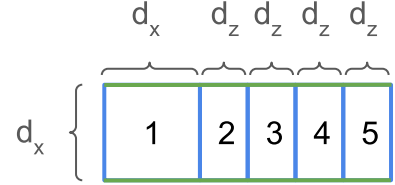}
\caption{Size of data qubits. Logical $Z$ ($X$) operators on the boundaries are green (blue) colored.} 
\label{fig:data-qubits-size}
\end{figure}

To assess the logical error rate of a distilled magic state, it is essential to evaluate errors introduced by each operation within the distillation process.
Let us begin with idle errors occurring on data qubits.
The distillation protocol consists of $\frac{\pi}{8}$ rotations, with each rotation's axis being commutative with $Z$ operators, allowing $Z$ errors occurring on qubits 2--5 to be detected by the final $X$ measurements.
In contrast, $Z$ errors on qubit 1 and $X$ errors across all the data qubits may elude detection, and hence they need to be suppressed more strongly.
As demonstrated in \autoref{fig:data-qubits-size}, this is achieved using two distinct code distances, namely $d_X$ and $d_Z$ where $d_X \geq d_Z$, to protect the data qubits against idle errors.

Next, let us consider errors caused by faulty $T^\dagger$ measurements.
A $\frac{\pi}{8}$ rotation along $P$ translates a Pauli error from its faulty $T^\dagger$ measurement into a rotation along $P$, which is detectable by the final $X$ measurements, similarly to $Z$ errors on qubits 2--5.
The error rate of the Pauli measurement $M_{PZ}$ in a faulty $T^\dagger$ measurement (\autoref{fig:faulty-t-measurement}) is determined by the number of cycles spent for the measurement, which we call $d_m$.
Given the observation, we set $d_m = d_Z$, and hence, each round completes in $d_Z$ cycles.

\begin{figure}[t!]
\centering
\includesvg[width=8cm]{fig/135-rotation-with-long-height.svg}
\caption{$\frac{\pi}{8}$ rotation along $Z_1Z_3Z_5$ (left) and logical operators on the boundary during the rotation (right).}
\label{fig:merged-state-error-analysis}
\end{figure}

Idle errors on separate data qubits are covered by the above discussion, but idle errors during $\frac{\pi}{8}$ rotations are not.
Consider, for instance, performing a $\frac{\pi}{8}$ rotation along $Z_1Z_3Z_5$, as illustrated in \autoref{fig:merged-state-error-analysis}~(left).
During the rotation, data qubits involved in the rotation and the logical qubit on which the faulty $T^\dagger$ gate is performed ($b$ in the figure) are merged through the ancilla region (the purple rectangle in the figure) with the lattice surgery.
\autoref{fig:merged-state-error-analysis}~(right) illustrates the logical operators on the boundary of the merged qubits.
Green (blue) lines represent logical $Z$ ($X$) operators.
As apparent in the figure, logical $Z$ operators on the boundary are distant from each other, which means the lengths of logical $X$ operators are long.
Hence, the error rate of idle $X$ errors during the rotation is very small.
In contrast, $Z$ errors can be problematic when $h$ in the figure is short.
\autoref{fig:merged-state-error-analysis}~(i), (ii), and (iii) represent $Z$ errors spanning $X$ operators on the boundary.
(i) is equivalent to $Z_1$ whereas (ii) is equivalent to $Z_1Z_3$ and (iii) is equivalent to $Z_1Z_3Z_b \simeq Z_5$.
In this case, (i) is the most severe because a $Z_1$ error is not detectable by the final $X$ measurements.
Therefore, we set $h = d_X$ to suppress the error with the same level of effectiveness as we do for idle $Z$ errors on qubit 1.
Let us call this distillation implementation (15-to-1)$_{d_X,d_Z}$ where $d_X$ and $d_Z$ are parameters. \footnote{The implementation in~\cite{Litinski2019magicstate} contains another distance parameter, $d_m$. For simplicity, in this study, we always set $d_m = d_Z$.}

\subsection{(15-to-1)$\times$(15-to-1) protocol implementation}
\label{subsec:preliminary-double-15-to-1}
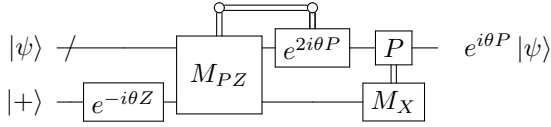
\begin{figure}[t!]
\centering
\[
\Qcircuit @C=.5em @R=.6em {
                      &         &                         & \controlo[1] \cwx[1]   & \cctrlo{1} \cwx[1]     & \\
  \lstick{\ket{\psi}} & {/} \qw & \qw                     & \multigate{1}{M_{PZ}}  & \gate{e^{2i{\theta}P}} & \gate{P}            & \qw & \rstick{e^{i{\theta}P}\ket{\psi}} \\
  \lstick{\ket{+}}    & \qw     & \gate{e^{-i{\theta}Z}}  & \ghost{M_{PZ}}         & \qw                    & \gate{M_X} \cwx[-1] &                                               \\
  }
\]
\caption{Circuit of $\theta$ rotation along P with state injection.}
\label{fig:state-injection}
\end{figure}

\begin{figure}[t!]
\centering
\includesvg[width=8.5cm]{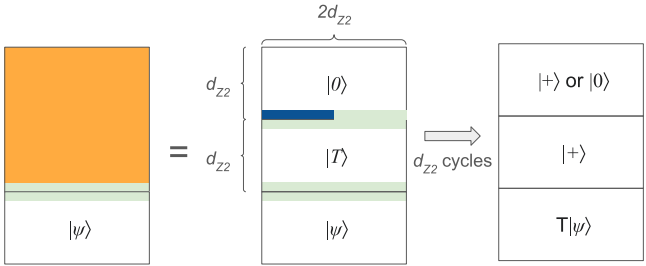}
\caption{Auto-correction block diagram.}
\label{fig:auto-correction}
\end{figure}

The (15-to-1)$_{d_{X1},d_{Z1}}\times$(15-to-1)$_{d_{X2},d_{Z2}}$ protocol implementation, which utilizes (15-to-1)$_{d_{X1},d_{Z1}}$ to generate first-level magic states, is similar to the single-level 15-to-1 protocol implementation, but there are a few differences.
The first difference is that the first-level distillation process operates at a slower pace compared to second-level $\frac{\pi}{8}$ rotations.
The implementation places multiple first-level distillation factories and establishes pipelines to offset the speed difference.
This also means that running more than two $\frac{\pi}{8}$ rotations within a round, which the single-level implementation does in the first and third rounds (\autoref{fig:single-level-15-to-1-implementation}), is difficult.
Hence the two-level distillation process requires 7.5 rounds to complete, which indicates that an eight-round distillation is followed by a seven-round distillation, rather than six rounds.
We also need to transfer magic states from the first-level distillation factories to perform second-level $\frac{\pi}{8}$ rotations; this transfer incurs an additional spatial cost.

Another difference is that we cannot use faulty $T^\dagger$ measurement at the second level.
Instead, we use state injection to perform a $\frac{\pi}{8}$ rotation, as illustrated in \autoref{fig:state-injection}.
This method requires a $\frac{\pi}{4}$ rotation with the probability of 1/2, potentially incurring a 50\% additional temporal overhead to the total distillation process, if unaddressed.
To eliminate the additional temporal overhead for the $\frac{\pi}{4}$ rotation, we use the auto-correction circuit.
While it uses more qubits (four $d_{Z2}{\times}d_{Z2}$ patches instead of one), we can perform a $\frac{\pi}{8}$ rotation with $d_{Z2}$ cycles, as depicted in \autoref{fig:auto-correction}.
For consistency in block diagrams, this is also denoted as an orange cell in block diagrams, because state injection and faulty $T^\dagger$ measurement serve analogous purposes and they are not used at the same time.
An auto-correction block can perform a $\frac{\pi}{8}$ rotation within $d_{Z2}$ cycles, but it requires another $d_{Z2}$ cycles to transfer a magic state into it.
Hence, to keep performing two $\frac{\pi}{8}$ rotations for each round, we need four auto-correction blocks in total.

\subsection{zero-level distillation}
\label{subsec:zero-level-distillation}
Zero-level distillation~\cite{Itogawa2024Efficient} is an efficient magic state distillation protocol which runs distillation on physical qubits rather than logical qubits.
It performs the Hadamard test of the logical $H$ gate on the Steane seven-qubit code~\cite{Steane1996ErrorCorrecting} to distill a magic state, and then the distilled magic state is teleported into the rotated surface code.
The original zero-level distillation generates $\ket{A} = e^{-\frac{i\pi}{8}Y}\ket{+}$, but it can also generate $\ket{T} = e^{-\frac{i\pi}{8}Z}\ket{+}$ by changing the rotation axis.

Zero-level distillation requires 40 physical qubits and 25 physical depths.
The 40 physical qubits fit into a single 5$\times$5 surface code patch.
One syndrome measurement on the rotated surface code consists of up to four CNOTs and two measurements.
Hence one cycle corresponds to six physical depths, and zero-level distillation completes in 5 cycles.

The logical error rate of a magic state generated by zero-level distillation is approximately $100p_{\mathrm{phys}}^2$ where $p_{\mathrm{phys}}$ is the physical error rate.
The failure rate is approximately $30\%$ when $p_{\mathrm{phys}} = 10^{-3}$ and $5\%$ when $p_{\mathrm{phys}} = 10^{-4}$.

While zero-level distillation offers preferable spatial and temporal overhead, it cannot directly generate magic states with a logical error rate smaller than $100p_{\mathrm{phys}}^2$.
Although it is possible to construct a two-level distillation protocol on top of zero-level distillation, similarly to the construction of the (15-to-1)$\times$(15-to-1) protocol discussed in \autoref{subsec:preliminary-double-15-to-1}, the spatial and temporal overhead of such a two-level distillation protocol, along with its impact on overall application performance, is not well understood.

In \autoref{sec:0+1-level distillation}, we will introduce (0+1)-level distillation, a two-level distillation protocol which combines zero-level distillation and the 15-to-1 distillation protocol, with its implementation.
In \autoref{sec:performance evaluation}, we will conduct performance analyses and compare its performance with conventional two-level distillation protocol implementations.

%% file: 3-main-proposal.tex
\section{(0+1)-level distillation}
\label{sec:0+1-level distillation}
This section introduces (0+1)-level distillation, a two-level distillation protocol that combines zero-level distillation and the 15-to-1 distillation protocol, and details its implementation.
The implementation builds upon the (15-to-1)$_{d_{X1},d_{Z1}}\times$(15-to-1)$_{d_{X2},d_{Z2}}$ implementation described in \autoref{subsec:preliminary-double-15-to-1}.
\autoref{subsec:0+1-distillation-trivial-integration} describes the integration of zero-level distillation with the 15-to-1 distillation protocol.
Following this, we outline strategies to refine the second-level 15-to-1 distillation protocol implementation for further efficiency (\autoref{subsec:0+1-distillation-teleportation} and \autoref{subsec:0+1-distillation-shortening-height-of-ancilla-region}).
Finally, in \autoref{subsec:0+1-distillation-implementation}, we combine these strategies and describe the (0+1)-level distillation implementation.
Throughout this section, we assume that $d_{X2} \geq d_{Z2} \geq 5$ to simplify our analysis.
With the assumption, zero-level distillation is performed in one $d_{Z2}{\times}d_{Z2}$ patch (spatially) and in $d_{Z2}$ cycles (temporally).

\subsection{Zero-level distillation integration}
\label{subsec:0+1-distillation-trivial-integration}
\begin{figure}[t!]
\centering
\includesvg[width=7cm]{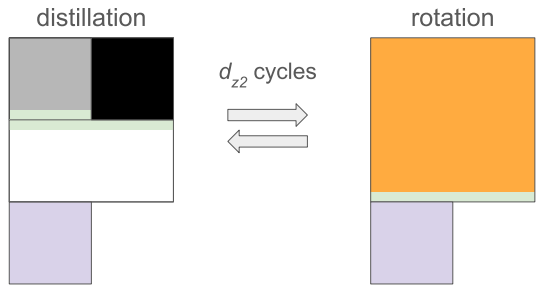}
\caption{Zero-level distillation in an auto-correction block. The gray area represents zero-level distillation. The black area represents unused qubits.}
\label{fig:auto-correction-and-zero-level-distillation}
\end{figure}

Zero-level distillation offers significant improvements in spatial and temporal efficiency.
By employing it to generate first-level magic states, we significantly reduce the spatial overhead associated with the two-level distillation.
With our assumption described above, zero-level distillation is performed in one $d_{Z2}{\times}d_{Z2}$ patch (spatially) and in $d_{Z2}$ cycles (temporally).
Hence, we can perform zero-level distillation in the auto-correction block, which eliminates the spatial cost for first-level distillation factories and qubits necessary to transfer the first-level magic states.
\autoref{fig:auto-correction-and-zero-level-distillation} visually presents this integration, showcasing both the distillation and $\frac{\pi}{8}$ rotation processes, where a zero-level distillation block is represented as a gray cell.

\subsection{Teleportation}
\label{subsec:0+1-distillation-teleportation}
\begin{figure}[t!]
\centering
\[
\Qcircuit @C=.5em @R=.6em { 
  \lstick{\ket{\psi}} & \qw & \multigate{1}{M_{ZZ}} & \qw       & \gate{M_X} \cwx[1] &                           \\
  \lstick{\ket{+}}    & \qw     & \ghost{M_{ZZ}}    & \gate{X}  & \gate{Z}           & \qw &\rstick{\ket{\psi}}  \\
                      &         & \control[-1] \cwx[-1]   & \cctrl{-1} \cwx[-1]      &                           \\
  }
\]
\caption{Teleportation circuit.}
\label{fig:teleportation circuit}
\end{figure}
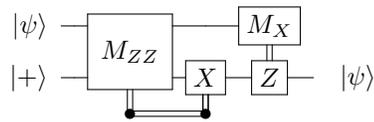

When the target state is $\ket{+}$, instead of performing a $\frac{\pi}{8}$ rotation on the target state, we can use quantum teleportation (\autoref{fig:teleportation circuit}).
Because teleportation does not require a $\frac{\pi}{4}$ fix-up rotation which is needed by state injection, we do not need an auto-correction block and hence it is more efficient.
Specifically, an auto-correction block uses four patches (spatially) and $2d_{Z2}$ cycles (temporally) whereas teleportation uses one patch (spatially) and $d_{Z2}$ cycles (temporally).

However, the efficiency offered by teleportation comes with caveats.
One is that, as stated above, teleportation can be used only when the target state is $\ket{+}$.
This also means that when using teleportation, the target qubit must not be involved in other rotations.
For instance, in the first round of the single-level 15-to-1 distillation (\autoref{fig:single-level-15-to-1-implementation}), rotations $Z_2$, $Z_3$, $Z_4$ and $Z_2Z_3Z_4$ are performed simultaneously.
This is not possible with teleportation.

Next, the desired state is $T^\dagger\ket{+} = e^{\frac{i\pi}{8}Z}\ket{+}$, rather than $\ket{T} = e^{-\frac{i\pi}{8}Z}\ket{+}$.
This is not problematic because zero-level distillation can generate the desired state, $e^{\frac{i\pi}{8}Z}\ket{+}$.

The last one is that zero-level distillation has a relatively high failure rate.
Faulty $T^\dagger$ measurement always succeeds although the operation itself is faulty.
Zero-level distillation, on the other hand, fails with a non-zero probability.
For example, with a physical error rate $p_\mathrm{phys}$ of $10^{-3}$, it fails at 30\%.
Consequently, when four teleportation operations are performed simultaneously to set qubits 2--5, at least one teleportation fails at 76\%, which results in a longer distillation process.

In \autoref{subsec:preliminary-single-15-to-1}, $X$ errors on qubits 2--5 are suppressed more strongly than $Z$ errors on them.
One may wonder if errors on the magic state lead to problems, especially when there is an $X$ error and $XT^\dagger\ket{+}$ is teleported to the target qubit.
This is not problematic, though, because $XT^\dagger\ket{+} = e^{-\frac{i\pi}{4}Z}T^\dagger\ket{+}$ (up to global phase), and the error is detectable by the final $X$ measurements.

\subsection{Shortening the height of the ancilla region}
\label{subsec:0+1-distillation-shortening-height-of-ancilla-region}
In \autoref{subsec:preliminary-single-15-to-1}, the height of the ancilla regions of the 15-to-1 implementation, which is $h$ in \autoref{fig:merged-state-error-analysis}, is set to $d_X$.
This setting is based on the observation that some $Z$ error chains of length $h$ are equivalent to a $Z$ error on qubit 1, which is an undetectable error.
However, not all $Z$ error chains pose the same level of threat;
for instance, $Z$ error chains on \autoref{fig:merged-state-error-analysis}~(ii) and (iii) are equivalent to $Z_1Z_3$ and $Z_5$, respectively, both of which are detectable.
In this subsection, we refine the layout of the second-level distillation process to avoid undetectable $Z$ error chains of length $h$, thereby allowing $h$ to be reduced, for further spatial efficiency.

\begin{figure}[t!]
\centering
\includesvg[width=6cm]{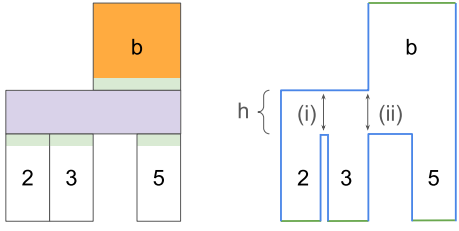}
\caption{$\frac{\pi}{8}$ rotation along $Z_2Z_3Z_5$ (left) and logical operators on the boundary during the rotation (right).}
\label{fig:235-rotation with shortened height}
\end{figure}

When qubit 1 is not involved in a $\frac{\pi}{8}$ rotation, no error chains on the ancilla region contain errors equivalent to a $Z_1$ error, thereby making all $Z$ errors detectable.
\autoref{fig:235-rotation with shortened height} illustrates one such example.
$Z$ error chains on \autoref{fig:235-rotation with shortened height}~(i) and (ii) are equivalent to $Z_2$ and $Z_2Z_3$, respectively.
Consequently, under this condition, $h$ can be reduced.

\begin{figure}[t!]
\centering
\includesvg[width=8.5cm]{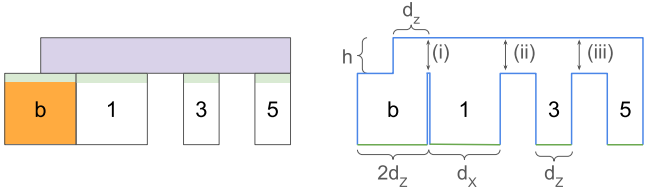}
\caption{$\frac{\pi}{8}$ rotation along $Z_1Z_3Z_5$ (left) and logical operators on the boundary during the rotation (right).}
\label{fig:135-rotation with shortened height}
\end{figure}

Even when qubit 1 is involved in a $\frac{\pi}{8}$ rotation, undetectable $Z$ errors can be circumvented by strategically placing auto-correction blocks.
The interpretation of a $Z$ error chain that connects the top and bottom sides of the ancilla region is determined by the arrangement of data qubits and the auto-correction block.
A $Z$ error chain on \autoref{fig:merged-state-error-analysis}~(i) is equivalent to $Z_1$ because it is topologically equivalent to a chain that connects the left and right edges of qubit 1.
This situation can be avoided by placing the auto-correction block immediately to the left of qubit 1.
\autoref{fig:135-rotation with shortened height} depicts the $\frac{\pi}{8}$ rotation along $Z_1Z_3Z_5$ with the layout.
$Z$ error chains on \autoref{fig:135-rotation with shortened height}~(i), (ii), and (iii) are equivalent to $Z_1Z_3Z_5$, $Z_3Z_5$, and $Z_5$, respectively, all of which are detectable.

Note that not all detectable errors are equally harmful.
The 15-to-1 distillation protocol, which performs 15 rotations, can detect up to two arbitrary rotation errors, contributing to its error suppression capability, characterized by $O(p^3)$.
However, certain idle errors occurring during a $\frac{\pi}{8}$ rotation compromise this property.
For instance, a $Z$ error chain on \autoref{fig:235-rotation with shortened height}~(ii) is equivalent to a $Z_2Z_3$ error.
While this error itself is detectable, the combination of $Z_2Z_3$ and $Z_1Z_2Z_3$ errors becomes undetectable, given that $Z_2Z_3 \cdot Z_1Z_2Z_3 = Z_1$.
Hence, a $Z_2Z_3$ error is more harmful than other errors, such as $Z_4$.
Therefore, when prioritizing magic state fidelity over distillation overhead, it is advisable to set $h$ larger than $d_{Z2}$.

\subsection{(0+1)-level distillation}
\label{subsec:0+1-distillation-implementation}
\begin{figure}[t!]
\centering
\includesvg[width=6cm]{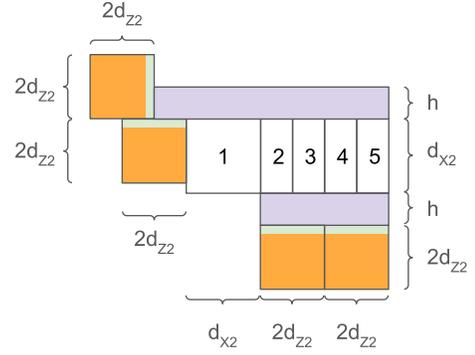}
\caption{(0+1)-level distillation layout.
         Numbered cells represent data qubits.
         Purple cells depicts the ancilla regions.
         Orange cells with green boundaries represent qubits used for zero-level distillation and auto-correction.}
\label{fig:0+1-distillation layout}
\end{figure}

In this subsection, we outline the implementation of (0+1)-level distillation, which has distance parameters $(d_{X2}, d_{Z2}, h)$.
It is designed to reduce the spatiotemporal overhead of the conventional (15-to-1)$\times$(15-to-1) distillation protocol, by integrating the strategies we discussed in previous subsections.
\autoref{fig:0+1-distillation layout} depicts the layout of the distillation circuit.
The top ancilla region is used for $\frac{\pi}{8}$ rotations involving qubit 1, while the bottom ancilla region is used for $\frac{\pi}{8}$ rotations that do not involve qubit 1.
As explored in \autoref{subsec:0+1-distillation-shortening-height-of-ancilla-region}, the parameter $h$, representing the height of both ancilla regions, can effectively be reduced to below $d_{X2}$, bringing improvements in spatial efficiency.

\begin{figure*}[t!]
\centering
\includesvg[width=18cm]{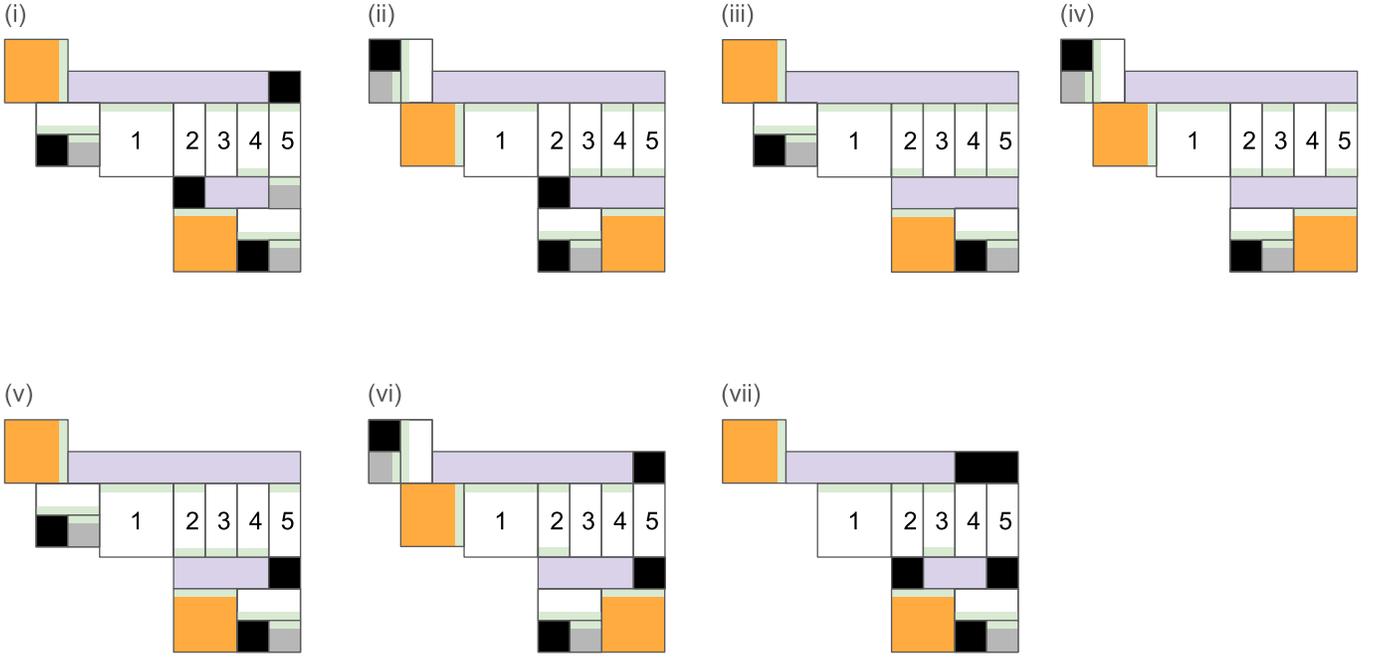}
\caption{(0+1)-level distillation process consisting of seven rounds.
         Numbered cells represent data qubits.
         Purple cells depict the ancilla regions.
         Black cells represent unused qubits.}
\label{fig:0+1-distillation process}
\end{figure*}

\autoref{fig:0+1-distillation process} illustrates the distillation process, which consists of seven rounds.
For example, in the first round, two rotations along $Z_1Z_3Z_4$ and $Z_4$, and one teleportation operation targeting qubit 5, are performed.
Given that the distillation process consists of seven rounds, the active and inactive auto-correction blocks alternate roles in the subsequent distillation; for example, the right-bottom auto-correction block is active in even-numbered rounds in the figure whereas it will be active in odd-numbered rounds in the next distillation.
The process illustrated in the figure is designed to seamlessly accommodate the role alternation;
for instance, the rotation along $Z_4$ can be performed with the right-bottom auto-correction block because the bottom ancilla region connects qubit 4 with the block.
At the end of the final round, qubits 2-5 are measured and the measurement results are checked to detect errors, while qubit 1 is reset.
Consequently, establishing entanglement between qubit 1 and an external logical qubit beforehand is essential for transferring the magic state out of the distillation factory.
Once qubit 1 is reset, the magic state is teleported to the entangled qubit, making it available for the consumer's use.

\begin{figure}[t!]
\centering
\includesvg[width=8cm]{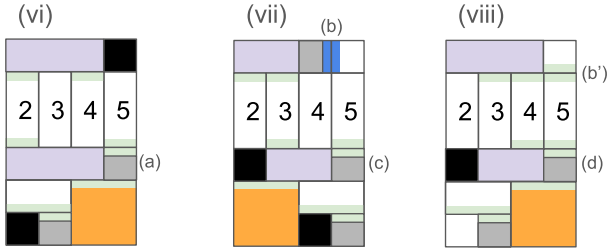}
\caption{Multiple distillation trials.
         (a)--(d) represent distillation operations.
         (b') is a teleportation operation that transfers the magic state generated by (b) to qubit 5.}
\label{fig:distillation failure mitigation}
\end{figure}

The distillation process performs two rotations for each round, but it performs an additional teleportation operation in the first round, which may incur an extra time cost because the failure probability of zero-level distillation $p_{\mathrm{fail}}$ is too large, as discussed in \autoref{subsec:0+1-distillation-teleportation}.
To mitigate the impact of these failures, we utilize unused qubits that are depicted as black-colored areas in \autoref{fig:0+1-distillation process}.
\autoref{fig:distillation failure mitigation} illustrates the mitigation approach through distillation rounds 6, 7, and 8, where round 8 is the first round of the next distillation.
For clarity, areas not related to qubit 5 are excluded from the figure.
Given that qubit 5 is not involved in $\frac{\pi}{8}$ rotations performed during rounds 6 and 7, it is measured at the end of round 5, freeing up space for the subsequent distillation.
The elements labeled (a), (b), (c), and (d) in the figure represent zero-level distillation instances, while (b') denotes a teleportation operation.
\autoref{alg:multiple distillation trials} details the execution of these operations.
Because one of distillation (a)--(d) needs to succeed to avoid retry, the probability of having to retry becomes $p_{\mathrm{fail}}^4$, which is sufficiently small in our settings.

\begin{figure}[!t]
\begin{algorithm}[H]
\caption{Execution of multiple distillation trials.}
\label{alg:multiple distillation trials}
\begin{algorithmic}[1]
  \STATE Perform (a) in round 6.
  \IF {(a) fails}
   \STATE Perform (b) in round 7.
   \STATE Perform (c) in round 7.

   \IF {(c) succeeds}
    \STATE Do nothing.
   \ELSIF {(b) succeeds}
    \STATE Perform (b') in round 8.
   \ELSE
     \STATE Perform (d) repeatedly until it succeeds in round 8.
   \ENDIF
  \ENDIF
\end{algorithmic}
\end{algorithm}
\end{figure}

%% file: 4-performance-evaluation.tex
\section{Performance evaluation}
\label{sec:performance evaluation}
In this section, we conduct a comprehensive performance evaluation of (0+1)-level distillation and compare its performance with conventional protocols.
Initially, the spatiotemporal overhead associated with these protocols is assessed, as detailed in \autoref{subsec:performance-evaluation-overhead}.
The analysis then progresses to explore how this performance impacts a critical application, namely Hamiltonian simulation, which is outlined in \autoref{subsec:performance-evaluation-application}.

\subsection{Spatiotemporal overhead}
\label{subsec:performance-evaluation-overhead}
To calculate the spatiotemporal overhead associated with distillation implementations, we employed Python and Mathematica scripts provided in~\cite{Litinski2019MagicStateScript}, which are originally designed for analyses in~\cite{Litinski2019magicstate}.
For our specific evaluation of (0+1)-level distillation, we extended the Python script's functionality.
This extension includes enhancing the calculation precision, enabling it to accommodate high-fidelity settings, which are crucial for ensuring the accuracy of our distillation overhead assessments.

\autoref{fig:spatiotemporal overhead with pphys = 10^{-4}} illustrates the spatiotemporal overhead associated with various distillation protocols at a physical error rate ($p_{\mathrm{phys}}$) of $10^{-4}$.
The figure showcases four different distillation protocols, each represented by multiple points or stars to reflect variations in specific parameters.
For instance, the single-level 15-to-1 protocol has two parameters $d_X$ and $d_Z$, and configurations such as (15-to-1)$_{7,3}$ and (15-to-1)$_{9,3}$ are each represented as individual blue points\footnote{As noted in \autoref{subsec:preliminary-single-15-to-1}, the single-level 15-to-1 protocol actually has three parameters, $d_X$, $d_Z$ and $d_m$. We include parameter settings where $d_Z \neq d_m$ in the figure.}.
The horizontal axis denotes the logical error rate of the distilled magic states, whereas the vertical axis quantifies the spatiotemporal overhead, calculated as the product of the number of physical qubits and the cycle count.
Notably, (0+1)-level distillation, which is shown as purple stars in the figure, significantly reduces the spatiotemporal overhead for generating magic states within an error rate range of $[5 \times 10^{-17}, 10^{-11}]$.
Specifically, it achieves a 70\% reduction in overhead for magic states with an error rate around $10^{-13}$, and a 65\% reduction for those around $10^{-15}$, compared to the (15-to-1)$\times$(15-to-1) protocol.

\begin{figure}[t!]
\centering
\includesvg[width=8.5cm]{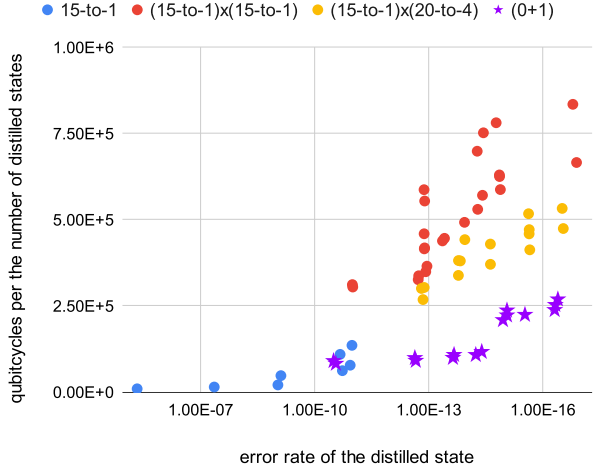}
\caption{Spatiotemporal overhead of distillation protocols with $p_{\mathrm{phys}} = 10^{-4}$.}
\label{fig:spatiotemporal overhead with pphys = 10^{-4}}
\end{figure}

The (15-to-1)$\times$(20-to-4) protocol exhibits preferable overhead compared to the (15-to-1)$\times$(15-to-1) protocol in this range, yet (0+1)-level distillation presents even lower overhead overall.
It is worth noting that the (15-to-1)$\times$(20-to-4) protocol, which generates four magic states, faces challenges in the timely transfer of distilled magic states from the factory, often requiring additional spatial resources, which is a difficulty not encountered with (0+1)-level distillation or the (15-to-1)$\times$(15-to-1) protocol.

\begin{figure}[t!]
\centering
\includesvg[width=8.5cm]{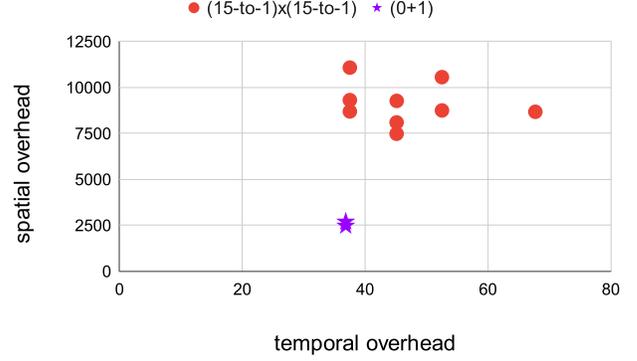}
\caption{Spatial and temporal overhead of the (15-to-1)$\times$(15-to-1) protocol and (0+1)-level distillation at error rates around $10^{-13}$.}
\label{fig:spatial-vs-time-reduction}
\end{figure}

One might question whether the reduced overhead of (0+1)-level distillation, as compared to the (15-to-1)$\times$(15-to-1) protocol, is more a result of reductions in spatial overhead or temporal overhead.
In many cases, the reduction in spatial overhead contributes more significantly than that in temporal overhead.
For example, \autoref{fig:spatial-vs-time-reduction} compares the spatial overhead (in physical qubits) and the temporal overhead (in cycles) for both protocols at error rates around $10^{-13}$.
This comparison demonstrates that the spatial overhead of (0+1)-level distillation is significantly lower than that of the (15-to-1)$\times$(15-to-1) protocol, while the differences in temporal overhead are less pronounced.

We observe a similar trend in \autoref{fig:spatiotemporal overhead with pphys = 10^{-3}}, which illustrates the spatiotemporal overhead associated with the distillation protocols at a physical error rate ($p_{\mathrm{phys}}$) of $10^{-3}$.
(0+1)-level distillation presents preferable overhead for generating magic states within an error rate range of $[5 \times 10^{-11}, 10^{-8}]$, demonstrating its consistent efficiency.

\begin{figure}[t!]
\centering
\includesvg[width=8.5cm]{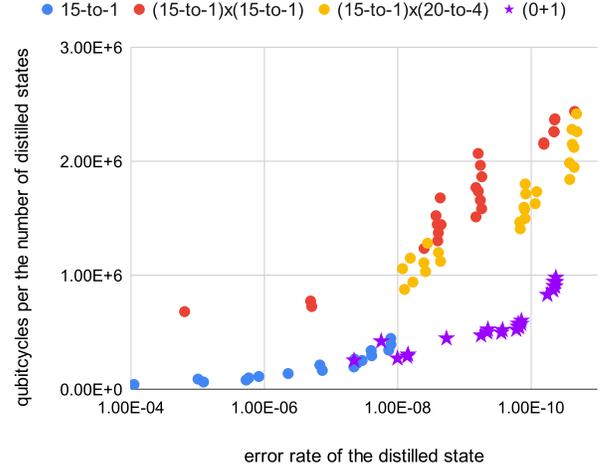}
\caption{Spatiotemporal overhead of distillation protocols with $p_{\mathrm{phys}} = 10^{-3}$.}
\label{fig:spatiotemporal overhead with pphys = 10^{-3}}
\end{figure}

\subsection{Case study for massively parallel magic state consumption}
\label{subsec:performance-evaluation-application}

In this subsection, we analyze the performance implication of the spatiotemporal reduction observed in \autoref{subsec:performance-evaluation-overhead} on quantum computation that consumes many magic states simultaneously.
Specifically, we analyze the performance of a Hamiltonian simulation with Trotterization.
Given a Hamiltonian $H = \sum_{j}P_j$ where each $P_j$ is a Pauli operator, one can compute the time evolution of the target system with the following approximation
\[
e^{-iHt} = (e^{\frac{-iHt}{M}})^M = (e^{\sum_j{\frac{-it}{M}}P_j})^M \approx (\prod_j{e^{\frac{-it}{M}P_j}})^M,
\]
where $M$ is a large integer.
$\prod_j{e^{\frac{-it}{M}P_j}}$ is called a \textit{Trotter step}, and $M$ is called the number of Trotter steps.
Note that the order of the Pauli operators in a Trotter step does not matter, given that
\[
e^{\frac{-it}{M}P_j}e^{\frac{-it}{M}P_k} \approx e^{\frac{-it}{M}(P_j + P_k)} = e^{\frac{-it}{M}(P_k + P_j)} \approx e^{\frac{-it}{M}P_k}e^{\frac{-it}{M}P_j}.
\]

\begin{figure}[t!]
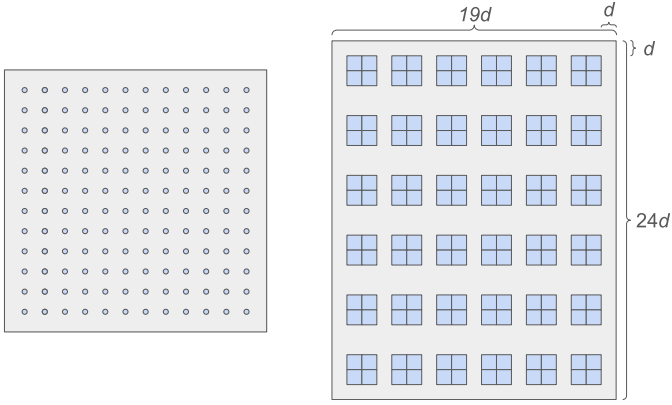

\begin{minipage}{3.5cm}
\centering
\includesvg[width=3.5cm]{fig/hamiltonian-simulation-target-system.svg}
\end{minipage}
\hfill
\begin{minipage}{4.5cm}
\centering
\includesvg[width=4.5cm]{fig/hamiltonian-simulation-layout.svg}
\end{minipage}
\caption{(left) Simulation target system.
         Each circle represents a site.
         (right) Qubit layout for the simulation.
         Each blue square is a logical data qubit corresponding to a site.
         Gray areas can be used to perform Pauli rotations.}
\label{fig:hamiltonian-simulation-target-system}
\end{figure}

To conduct a thorough performance analysis of this computation, detailed information of the Hamiltonian $H = \sum_{j}P_j$ is required.
In this subsection, let us assume that the target system consists of 12$\times$12 = 144 square-shaped sites (\autoref{fig:hamiltonian-simulation-target-system}~(left)), and $H$ is the sum of Pauli operators acting on every neighboring two sites.
This means $H$ is the sum of 12$\times$11$\times$2 = 264 weight-two Pauli operators.
\autoref{fig:hamiltonian-simulation-target-system}~(right) shows our arrangement of the logical qubits for the simulation.
Let us call it the \textit{central arena} because we will place magic state distillation factories around it.
It consists of $24\times19 = 456$ $d{\times}d$ patches where $d$ is the code distance.

Each $e^{\frac{-it}{M}P_j}$ is a Pauli rotation along $P_j$.
With an efficient gate decomposition method~\cite{Kliuchnikov2023Shorter}, the number of $\pm\frac{\pi}{8}$ rotations required to implement $e^{\frac{-it}{M}P_j}$ is approximately $(-0.53 \log_2\delta + 5)$ where $\delta$ is the precision of the rotation angle.
Hence, the total number of magic states required in the simulation is approximately $(-140\log_2\delta + 1320)M$.

\begin{figure}[t!]
\centering
\includesvg[width=8cm]{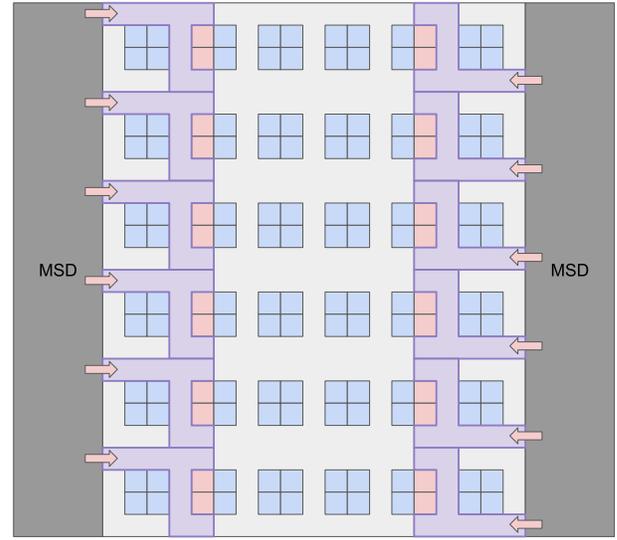}
\caption{Qubit layout for the simulation with magic state distillation factories and twelve Pauli rotations performed simultaneously.
         Dark gray areas represent qubits used to distill magic states.
         Pink squares represent logical data qubits involved in Pauli rotations.
         Each purple area represents qubits used to perform the corresponding rotation.
         Each pink arrow represents a flow of a magic state used for the corresponding rotation.}
\label{fig:hamiltonian-simulation-layout-with-msd}
\end{figure}

In theory, Pauli rotations acting on different qubits can be performed simultaneously, allowing for up to 72 rotations to be performed in parallel.
In practice, this is not possible even with sufficient magic states, because the layout in \autoref{fig:hamiltonian-simulation-target-system}~(right) cannot provide the magic states to all the rotations simultaneously.
As illustrated in \autoref{fig:hamiltonian-simulation-layout-with-msd}, it is possible to perform twelve rotations simultaneously, by providing six magic states from the left side and six magic states from the right side.
With a magic state, a $\pm\frac{\pi}{8}$ rotation can be performed in $d$ cycles, and hence, the total time cost of the computation is approximately $\frac{-140\log_2\delta + 1320}{12}Md$ cycles.
Note that the time cost of Clifford gates in the decomposed Pauli rotations is negligible with the layout.

Now we are ready to set concrete parameters, determine the code distance, and choose appropriate parameters for magic state distillation protocols.
Let us set $M = 10^6$, $\delta = 10^{-11}$, and $p_{\mathrm{phys}} = 10^{-4}$.
We also assume that we want the error rate of Clifford gates and non-Clifford gates to be below 0.5\%, respectively.
Using the surface code error rate approximation~\cite{Litinski2019GameOfSurfaceCodes, Gidney2019EfficientMagicState}
\[
p_L(d) \coloneqq 0.1(100p_{\mathrm{phys}})^{\frac{d + 1}{2}},
\]
we have $456 \times \frac{-140\log_2\delta + 1320}{12}Md \times p_L(d) < 0.005$, and $d = 13$ is the smallest odd integer satisfying this inequality.
The error rate of a magic state needs to be below $\frac{0.005}{-140M\log_2\delta + 1320M} \approx 7.8 \times 10^{-13}$.
Hence, we use the magic state distillation factories listed in \autoref{tab:distillation-protocols-for-hamiltonian-simulation}.
The rightmost column in the table represents the number of distillation factories required to provide sufficient magic states.

\begin{table}[t!]
  \centering
  \caption{Magic state distillation factories used for the simulation.}
  \label{tab:distillation-protocols-for-hamiltonian-simulation}
  \begin{tabular}{|c|r|r|r|r|}
    \hline
    \textbf{protocol} & \textbf{time} & \textbf{space} & \textbf{error rate} & \textbf{\#$_\mathrm{MSD}$} \\
    \hline
    \hline
      (15-to-1)$\times$(15-to-1) & 38 &  8690 & $1.9 \times 10^{-13}$ & 36 \\
    \hline
      (15-to-1)$\times$(20-to-4) & 70 & 15328 & $1.4 \times 10^{-13}$ & 18 \\
    \hline
      (0+1)                      & 37 &  2470 & $2.2 \times 10^{-13}$ & 36 \\
    \hline
  \end{tabular}
\end{table}

The spatial cost of the simulation, which is the total number of physical qubits required for the simulation, is the sum of the spatial cost of the central arena and the spatial cost of magic state distillation factories.
The spatial cost of a magic state distillation factory is the spatial cost listed in \autoref{tab:distillation-protocols-for-hamiltonian-simulation}.
We also allocate one logical qubit for each (15-to-1)$\times$(15-to-1) and (0+1)-level distillation factory, and four logical qubits for each (15-to-1)$\times$(20-to-4) factory, to store distilled magic states.
In the distillation area in \autoref{fig:hamiltonian-simulation-layout-with-msd}, some qubits are used to transfer magic states, but for simplicity of the analysis, we ignore this spatial cost. 

The spatial cost of the simulation using the (15-to-1)$\times$(15-to-1) protocol is $4.8 \times 10^5$, whereas for the (15-to-1)$\times$(20-to-4) protocol, it is $4.5 \times 10^5$, and for (0+1)-level distillation, the cost is $2.6 \times 10^5$.
Therefore, in this case, adopting (0+1)-level distillation reduces the total spatial cost by 47\% compared to using the (15-to-1)$\times$(15-to-1) protocol, and by 44\% compared to using the (15-to-1)$\times$(20-to-4) protocol.
This significant reduction highlights the impact of efficiency of (0+1)-level distillation in computation that consumes many magic states simultaneously.

In \autoref{tab:distillation-protocols-for-hamiltonian-simulation}, (0+1)-level distillation offers 72\% distillation spatiotemporal overhead reduction compared to the (15-to-1)$\times$(15-to-1) protocol, whereas it offers \textit{only} 47\% total spatial cost reduction.
This is because the central arena (\autoref{fig:hamiltonian-simulation-target-system} (right)) is not affected by the distillation protocol change.
In other words, the spatial cost of magic state distillation factories relative to the total spatial cost limits the total spatial cost reduction.

\autoref{fig:hamiltonian-simulation-layout-with-msd} consumes twelve magic states simultaneously to reduce the time cost of the computation.
We can further reduce the time cost by computing $e^{-i{\theta}Z}\ket{+}$ using magic states and gate decomposition in the dark gray areas in \autoref{fig:hamiltonian-simulation-layout-with-msd}, and use state injection (\autoref{fig:state-injection}) to perform a Pauli rotation $e^{i{\theta}P}$.
Although the number of required magic states is doubled on average, this configuration offers a much shorter execution time than the one discussed above, with sufficient magic states: the time cost is $\frac{264 \times 2}{12}Md$ cycles rather than $\frac{-140\log_2\delta + 1320}{12}Md$ cycles.
Because this configuration uses more magic states in a shorter period of time, it requires much more distillation factories and the total spatial cost reduction offered by (0+1)-level distillation would be more significant.

Let us also make an estimation with a larger physical error rate and a smaller number of Trotter steps.
We set $M = 10^4$, $\delta = 10^{-9}$, and $p_{\mathrm{phys}} = 10^{-3}$, which leads us to set $d = 23$.
The error rate of a magic state needs to be below $\frac{0.005}{-140M\log_2\delta + 1320M} \approx 9.1 \times 10^{-11}$, and we use magic state distillation factories listed in \autoref{tab:distillation-protocols-for-hamiltonian-simulation-10_3}. 

\begin{table}[t!]
  \centering
  \caption{Magic state distillation factories used for the simulation, $p_{\mathrm{phys}} = 10^{-3}$}
  \label{tab:distillation-protocols-for-hamiltonian-simulation-10_3}
  \begin{tabular}{|c|r|r|r|r|}
    \hline
    \textbf{protocol} & \textbf{time} & \textbf{space} & \textbf{error rate} & \textbf{\#$_\mathrm{MSD}$} \\
    \hline
    \hline
      (15-to-1)$\times$(15-to-1) &  92 & 23408 & $6.4 \times 10^{-11}$ & 48 \\
    \hline
      (15-to-1)$\times$(20-to-4) & 130 & 50116 & $8.6 \times 10^{-11}$ & 18 \\
    \hline
      (0+1)                      &  78 & 10614 & $5.7 \times 10^{-11}$ & 42 \\
    \hline
  \end{tabular}
\end{table}

The spatial cost of the simulation using the (15-to-1)$\times$(15-to-1) protocol is $1.7 \times 10^6$, whereas for the (15-to-1)$\times$(20-to-4) protocol, it is $1.5 \times 10^6$, and for (0+1)-level distillation, the cost is $9.7 \times 10^5$.
Therefore, in this setting, adopting (0+1)-level distillation reduces the total spatial cost by 41\% compared to using the (15-to-1)$\times$(15-to-1) protocol, and by 33\% compared to the (15-to-1)$\times$(20-to-4) protocol.

%% file: 5-conclusion.tex
\section{Conclusion}
\label{sec:conclusion}
In this study, we evaluated the spatial and temporal overhead of two-level distillation implementations generating relatively high-fidelity magic states, including ones incorporating zero-level distillation.
To this end, we introduced (0+1)-level distillation, a two-level distillation protocol which combines zero-level distillation and the 15-to-1 distillation protocol, with refining the second-level 15-to-1 distillation implementation to capitalize on the small footprint of zero-level distillation.

The numerical simulation in \autoref{subsec:performance-evaluation-overhead} shows that, under conditions of a physical error probability of $p_{\mathrm{phys}} = 10^{-4}$ ($10^{-3}$) and targeting an error rate for the magic state within $[5 \times 10^{-17}, 10^{-11}]$ ($[5 \times 10^{-11}, 10^{-8}]$), (0+1)-level distillation reduces the spatiotemporal overhead by more than 63\% (61\%) compared to the (15-to-1)$\times$(15-to-1) protocol and more than 43\% (44\%) compared to the (15-to-1)$\times$(20-to-4) protocol, offering a substantial efficiency gain over the traditional protocols.

The resource estimation of a Hamiltonian simulation in \autoref{subsec:performance-evaluation-application} shows that, with $p_{\mathrm{phys}} = 10^{-4}$ $(10^{-3})$, adopting (0+1)-level distillation reduces the total spatial cost by 47\% (41\%) compared to using the (15-to-1)$\times$(15-to-1) protocol, and by 44\% (33\%) compared to the (15-to-1)$\times$(20-to-4) protocol.
Furthermore, it is possible to consume more magic states simultaneously to reduce the time cost of the computation.
With such configurations, (0+1)-level distillation would offer even more significant total spatial cost reductions.

%% file: main.bbl
\begin{thebibliography}{10}
\providecommand{\url}[1]{#1}
\csname url@samestyle\endcsname
\providecommand{\newblock}{\relax}
\providecommand{\bibinfo}[2]{#2}
\providecommand{\BIBentrySTDinterwordspacing}{\spaceskip=0pt\relax}
\providecommand{\BIBentryALTinterwordstretchfactor}{4}
\providecommand{\BIBentryALTinterwordspacing}{\spaceskip=\fontdimen2\font plus
\BIBentryALTinterwordstretchfactor\fontdimen3\font minus \fontdimen4\font\relax}
\providecommand{\BIBforeignlanguage}[2]{{%
\expandafter\ifx\csname l@#1\endcsname\relax
\typeout{** WARNING: IEEEtran.bst: No hyphenation pattern has been}%
\typeout{** loaded for the language `#1'. Using the pattern for}%
\typeout{** the default language instead.}%
\else
\language=\csname l@#1\endcsname
\fi
#2}}
\providecommand{\BIBdecl}{\relax}
\BIBdecl

\bibitem{Shor1994Factoring}
P.~W. Shor, ``Algorithms for quantum computation: discrete logarithms and factoring,'' \emph{Proceedings 35th Annual Symposium on Foundations of Computer Science}, pp. 124--134, 1994.

\bibitem{aspuru2005simulated}
A.~Aspuru-Guzik, A.~D. Dutoi, P.~J. Love, and M.~Head-Gordon, ``Simulated quantum computation of molecular energies,'' \emph{Science}, vol. 309, no. 5741, pp. 1704--1707, 2005.

\bibitem{harrow2009quantum}
A.~W. Harrow, A.~Hassidim, and S.~Lloyd, ``Quantum algorithm for linear systems of equations,'' \emph{Physical review letters}, vol. 103, no.~15, p. 150502, 2009.

\bibitem{preskill2018quantum}
J.~Preskill, ``Quantum computing in the nisq era and beyond,'' \emph{Quantum}, vol.~2, p.~79, 2018.

\bibitem{fujii2015quantum}
K.~Fujii, \emph{Quantum Computation with Topological Codes: from qubit to topological fault-tolerance}.\hskip 1em plus 0.5em minus 0.4em\relax Springer, 2015, vol.~8.

\bibitem{Kitaev2003}
\BIBentryALTinterwordspacing
A.~Kitaev, ``Fault-tolerant quantum computation by anyons,'' \emph{Annals of Physics}, vol. 303, no.~1, p. 2–30, Jan. 2003. [Online]. Available: \url{http://dx.doi.org/10.1016/S0003-4916(02)00018-0}
\BIBentrySTDinterwordspacing

\bibitem{Bravyi1998}
S.~B. Bravyi and A.~Y. Kitaev, ``Quantum codes on a lattice with boundary,'' 1998.

\bibitem{Horsman2012}
\BIBentryALTinterwordspacing
D.~Horsman, A.~G. Fowler, S.~Devitt, and R.~V. Meter, ``Surface code quantum computing by lattice surgery,'' \emph{New Journal of Physics}, vol.~14, no.~12, p. 123011, dec 2012. [Online]. Available: \url{https://dx.doi.org/10.1088/1367-2630/14/12/123011}
\BIBentrySTDinterwordspacing

\bibitem{Litinski2019GameOfSurfaceCodes}
\BIBentryALTinterwordspacing
D.~Litinski, ``A {G}ame of {S}urface {C}odes: {L}arge-{S}cale {Q}uantum {C}omputing with {L}attice {S}urgery,'' \emph{{Quantum}}, vol.~3, p. 128, Mar. 2019. [Online]. Available: \url{https://doi.org/10.22331/q-2019-03-05-128}
\BIBentrySTDinterwordspacing

\bibitem{Bravyi2005}
\BIBentryALTinterwordspacing
S.~Bravyi and A.~Kitaev, ``Universal quantum computation with ideal clifford gates and noisy ancillas,'' \emph{Phys. Rev. A}, vol.~71, p. 022316, Feb 2005. [Online]. Available: \url{https://link.aps.org/doi/10.1103/PhysRevA.71.022316}
\BIBentrySTDinterwordspacing

\bibitem{Campbell2017Unified}
\BIBentryALTinterwordspacing
E.~T. Campbell and M.~Howard, ``Unified framework for magic state distillation and multiqubit gate synthesis with reduced resource cost,'' \emph{Phys. Rev. A}, vol.~95, p. 022316, Feb 2017. [Online]. Available: \url{https://link.aps.org/doi/10.1103/PhysRevA.95.022316}
\BIBentrySTDinterwordspacing

\bibitem{Gidney2019EfficientMagicState}
\BIBentryALTinterwordspacing
C.~Gidney and A.~G. Fowler, ``Efficient magic state factories with a catalyzed {$|CCZ\rangle$} to {$2|T\rangle$} transformation,'' \emph{{Quantum}}, vol.~3, p. 135, Apr. 2019. [Online]. Available: \url{https://doi.org/10.22331/q-2019-04-30-135}
\BIBentrySTDinterwordspacing

\bibitem{Litinski2019magicstate}
\BIBentryALTinterwordspacing
D.~Litinski, ``Magic {S}tate {D}istillation: {N}ot as {C}ostly as {Y}ou {T}hink,'' \emph{{Quantum}}, vol.~3, p. 205, Dec. 2019. [Online]. Available: \url{https://doi.org/10.22331/q-2019-12-02-205}
\BIBentrySTDinterwordspacing

\bibitem{Herr2017LatticeSurgery}
\BIBentryALTinterwordspacing
D.~Herr, F.~Nori, and S.~J. Devitt, ``Lattice surgery translation for quantum computation,'' \emph{New Journal of Physics}, vol.~19, no.~1, p. 013034, jan 2017. [Online]. Available: \url{https://dx.doi.org/10.1088/1367-2630/aa5709}
\BIBentrySTDinterwordspacing

\bibitem{Itogawa2024Efficient}
T.~Itogawa, Y.~Takada, Y.~Hirano, and K.~Fujii, ``Even more efficient magic state distillation by zero-level distillation,'' 2024.

\bibitem{Steane1996ErrorCorrecting}
\BIBentryALTinterwordspacing
A.~M. Steane, ``Error correcting codes in quantum theory,'' \emph{Phys. Rev. Lett.}, vol.~77, pp. 793--797, Jul 1996. [Online]. Available: \url{https://link.aps.org/doi/10.1103/PhysRevLett.77.793}
\BIBentrySTDinterwordspacing

\bibitem{Litinski2019MagicStateScript}
\BIBentryALTinterwordspacing
D.~Litinski, ``Resource-cost estimates for magic state distillation.'' [Online]. Available: \url{https://github.com/litinski/magicstates}
\BIBentrySTDinterwordspacing

\bibitem{Kliuchnikov2023Shorter}
\BIBentryALTinterwordspacing
V.~Kliuchnikov, K.~Lauter, R.~Minko, A.~Paetznick, and C.~Petit, ``Shorter quantum circuits via single-qubit gate approximation,'' \emph{Quantum}, vol.~7, p. 1208, Dec. 2023. [Online]. Available: \url{http://dx.doi.org/10.22331/q-2023-12-18-1208}
\BIBentrySTDinterwordspacing

\end{thebibliography}
